# Photoacoustic imaging beyond the acoustic diffraction-limit with dynamic speckle illumination and sparse joint support recovery

**ELIEL HOJMAN,[1] THOMAS CHAIGNE,[2,3,4] OREN SOLOMON,[5] SYLVAIN GIGAN,[3] EMMANUEL BOSSY,[2,6,7] YONINA C. ELDAR,[5] AND ORI KATZ[1,*]**

*[1]Department of Applied Physics, The Selim and Rachel Benin School of Computer Science & Engineering, The Hebrew University of Jerusalem, Jerusalem 9190401, Israel*
*[2] ESPCI ParisTech, PSL Research University, CNRS UMR 7587, INSERM U979, Institut Langevin, 1 rue Jussieu, 75005 Paris, France*
*[3] Laboratoire Kastler Brossel, Université Pierre et Marie Curie, Ecole Normale Supérieure, Collège de France, CNRS UMR 8552, 24 rue Lhomond,*
*75005 Paris, France*
*[4] Current affiliation: Exzellenzcluster NeuroCure, Charité Berlin, Humboldt University, Charitéstr. 1, 10117 Berlin, Germany*
*[5] Electrical Engineering Department, Technion, Haifa 32298, Israel.*
*[6] Optics Laboratory and Laboratory of Applied Photonics Devices, School of Engineering, EPFL, 1015 Lausanne, Switzerland*
*[7] Univ. Grenoble Alpes, CNRS, LIPHY, F-38000 Grenoble, France*

**orik@mail.huji.ac.il*

**Abstract:** In deep tissue photoacoustic imaging the spatial resolution is inherently limited by the acoustic wavelength. Recently, it was demonstrated that it is possible to surpass the acoustic diffraction limit by analyzing fluctuations in a set of photoacoustic images obtained under unknown speckle illumination patterns. Here, we purpose an approach to boost reconstruction fidelity and resolution, while reducing the number of acquired images by utilizing a compressed sensing computational reconstruction framework. The approach takes into account prior knowledge of the system response and sparsity of the target structure. We provide proof of principle experiments of the approach and demonstrate that improved performance is obtained when both speckle fluctuations and object priors are used. We numerically study the expected performance as a function of the measurement's signal to noise ratio and sample spatial-sparsity. The presented reconstruction framework can be applied to analyze existing photoacoustic experimental datasets containing dynamic fluctuations.

## References and links

1. V. Ntziachristos, "Going deeper than microscopy: the optical imaging frontier in biology," Nat Meth **7**, 603-614 (2010).
2. L. V. Wang and S. Hu, "Photoacoustic Tomography: In Vivo Imaging from Organelles to Organs," Science **335**, 1458-1462 (2012).
3. P. Beard, "Biomedical photoacoustic imaging," Interface focus **1**, 602-631 (2011).
4. T. Chaigne, J. Gateau, M. Allain, O. Katz, S. Gigan, A. Sentenac, and E. Bossy, "Super-resolution photoacoustic fluctuation imaging with multiple speckle illumination," Optica **3**, 54-57 (2016).
5. T. Dertinger, R. Colyer, G. Iyer, S. Weiss, and J. Enderlein, "Fast, background-free, 3D super-resolution optical fluctuation imaging (SOFI)," Proceedings of the National Academy of Sciences **106**, 22287-22292 (2009).
6. J. Gateau, T. Chaigne, O. Katz, S. Gigan, and E. Bossy, "Improving visibility in photoacoustic imaging using dynamic speckle illumination," Optics Letters **38**, 5188-5191 (2013).
7. MudryE, BelkebirK, GirardJ, SavatierJ, E. Le Moal, NicolettiC, AllainM, and SentenacA, "Structured illumination microscopy using unknown speckle patterns," Nat Photon **6**, 312-315 (2012).

8. J. Min, J. Jang, D. Keum, S.-W. Ryu, C. Choi, K.-H. Jeong, and J. C. Ye, "Fluorescent microscopy beyond diffraction limits using speckle illumination and joint support recovery," Scientific reports **3**(2013).
9. A. Liutkus, D. Martina, S. Popoff, G. Chardon, O. Katz, G. Lerosey, S. Gigan, L. Daudet, and I. Carron, "Imaging With Nature: Compressive Imaging Using a Multiply Scattering Medium," Sci. Rep. **4**(2014).
10. S. F. Cotter, B. D. Rao, K. Engan, and K. Kreutz-Delgado, "Sparse solutions to linear inverse problems with multiple measurement vectors," IEEE Transactions on Signal Processing **53**, 2477-2488 (2005).
11. S. Gazit, A. Szameit, Y. C. Eldar, and M. Segev, "Super-resolution and reconstruction of sparse sub-wavelength images," Optics Express **17**, 23920-23946 (2009).
12. A. Szameit, Y. Shechtman, E. Osherovich, E. Bullkich, P. Sidorenko, H. Dana, S. Steiner, E. B. Kley, S. Gazit, and T. Cohen-Hyams, "Sparsity-based single-shot subwavelength coherent diffractive imaging," Nature materials **11**, 455-459 (2012).
13. D. L. Donoho, "Compressed sensing," IEEE Transactions on information theory **52**, 1289-1306 (2006).
14. E. J. Candes and M. B. Wakin, "An Introduction To Compressive Sampling," Signal Processing Magazine, IEEE **25**, 21-30 (2008).
15. Y. C. Eldar, *Sampling Theory: Beyond Bandlimited Systems* (Cambridge University Press, 2015).
16. L. Zhu, W. Zhang, D. Elnatan, and B. Huang, "Faster STORM using compressed sensing," Nat Meth **9**, 721-723 (2012).
17. O. Katz, Y. Bromberg, and Y. Silberberg, "Compressive ghost imaging," Applied Physics Letters **95**, 131110 (2009).
18. J.-E. Oh, Y.-W. Cho, G. Scarcelli, and Y.-H. Kim, "Sub-Rayleigh imaging via speckle illumination," Optics Letters **38**, 682-684 (2013).
19. C. Ventalon and J. Mertz, "Quasi-confocal fluorescence sectioning with dynamic speckle illumination," Optics letters **30**, 3350-3352 (2005).
20. D. Lim, K. K. Chu, and J. Mertz, "Wide-field fluorescence sectioning with hybrid speckle and uniform-illumination microscopy," Optics letters **33**, 1819-1821 (2008).
21. M. Kim, C. Park, C. Rodriguez, Y. Park, and Y.-H. Cho, "Superresolution imaging with optical fluctuation using speckle patterns illumination," Scientific reports **5**(2015).
22. J. García, Z. Zalevsky, and D. Fixler, "Synthetic aperture superresolution by speckle pattern projection," Optics express **13**, 6073-6078 (2005).
23. M. G. Gustafsson, "Nonlinear structured-illumination microscopy: wide-field fluorescence imaging with theoretically unlimited resolution," Proceedings of the National Academy of Sciences of the United States of America **102**, 13081-13086 (2005).
24. E. Betzig, G. H. Patterson, R. Sougrat, O. W. Lindwasser, S. Olenych, J. S. Bonifacino, M. W. Davidson, J. Lippincott-Schwartz, and H. F. Hess, "Imaging Intracellular Fluorescent Proteins at Nanometer Resolution," Science **313**, 1642-1645 (2006).
25. M. J. Rust, M. Bates, and X. Zhuang, "Sub-diffraction-limit imaging by stochastic optical reconstruction microscopy (STORM)," Nat Meth **3**, 793-796 (2006).
26. M. Mishali and Y. C. Eldar, "Reduce and boost: Recovering arbitrary sets of jointly sparse vectors," IEEE Transactions on Signal Processing **56**, 4692-4702 (2008).
27. D. P. Wipf and B. D. Rao, "An empirical Bayesian strategy for solving the simultaneous sparse approximation problem," IEEE Transactions on Signal Processing **55**, 3704-3716 (2007).
28. Z. Zhang and B. D. Rao, "Sparse signal recovery with temporally correlated source vectors using sparse Bayesian learning," IEEE Journal of Selected Topics in Signal Processing **5**, 912-926 (2011).
29. M. Xu and L. V. Wang, "Universal back-projection algorithm for photoacoustic computed tomography," Physical Review E **71**, 016706 (2005).
30. J. W. Goodman, *Speckle phenomena in optics : theory and applications* (Roberts & Co., Englewood, Colo., 2007), pp. xvi, 387 p.
31. G. Ayers and J. C. Dainty, "Iterative blind deconvolution method and its applications," Optics letters **13**, 547-549 (1988).
32. S. Gleichman and Y. C. Eldar, "Blind Compressed Sensing," IEEE Transactions on Information Theory **57**, 6958-6975 (2011).
33. T. W. Murray, M. Haltmeier, T. Berer, E. Leiss-Holzinger, and P. Burgholzer, "Super-resolution photoacoustic microscopy using blind structured illumination," Optica **4**, 17-22 (2017).
34. E. Hojman, "Photoacoustic object recovery through M-SBL [Data set]," Zenodo https://doi.org/10.5281/zenodo.268701 (2017).

## 1. Introduction

Optical microscopy is an invaluable tool in biomedical investigation and clinical diagnostics. However, its use is limited to depths of not more than a fraction of a millimeter inside tissue due to light scattering. At depths beyond a few hundred microns light scattering in tissue prevents the ability to focus light to its diffraction limit. While non-optical imaging techniques, employing non-ionizing radiation such as ultrasound, allow deeper investigations, they typically possess inferior resolution and generally do not permit microscopic studies of cellular structures at depths of more than a millimeter [1]. One of the leading approaches for deep tissue optical imaging is photoacoustic imaging/tomography [2, 3]. Photoacoustic imaging relies on the generation of ultrasonic waves by absorption of light in a target structure under pulsed optical illumination. Ultrasonic waves are produced via thermo-elastic stress generation, and propagate to an externally placed ultrasonic detector-array without being scattered. Photoacoustic imaging thus provides images of optical contrast with a spatial resolution limited by acoustic diffraction. Ultimately, the ultrasound resolution in soft tissue is limited by the attenuation of high frequency ultrasonic waves. As a result, the depth-to-resolution ratio of deep-tissue photoacoustic imaging is ~100-200 in practice [2, 3]. For example, at a depth of 5 mm one can expect a resolution of around 20 μm at best, more than an order of magnitude above the optical diffraction limit.

Recently, it has been demonstrated that the conventional acoustic diffraction-limit can be overcome by exploiting temporal fluctuations in photoacoustic signals originating from illuminating the sample with dynamically varying optical speckle patterns [4]. This work was inspired by the notion of super-resolution optical fluctuation imaging (SOFI) [5], developed for fluorescence microscopy. In SOFI, a resolution beyond the diffraction limit is obtained via high-order statistical analysis of temporal fluctuations of fluorescence, recorded in a sequence of images. To apply SOFI in photoacoustics a set of random, unknown optical speckle illumination patterns was used as a source of fluctuations for super-resolution photoacoustic imaging [4] [Fig.1(a-b)]. Using this approach, an effective resolution enhancement of ~1.6 beyond the acoustic diffraction-limit was obtained by analyzing the temporal fluctuations' second moment (variance) using a set of 100 photoacoustic images. Obtaining higher resolution by analyzing higher statistical moments with such a limited number of images results in strong artifacts due to statistical noise, caused by the insufficient number of analyzed frames [Fig. 1(b), right inset].

Here, we show that by adapting an advanced computational reconstruction algorithm based on a compressed-sensing framework it is possible to obtain an enhancement in resolution and reconstruction fidelity in photoacoustic imaging beyond that possible with the basic statistical fluctuation analysis of SOFI [4], while using the same experimentally obtained dataset [Fig. 1(c)]. Specifically, we recognize that photoacoustic imaging under dynamic unknown speckle illumination [4, 6] is an instance of blind structured illumination microscopy (blind-SIM) [7, 8]. Since the photoacoustic signal generation and detection is a linear process, reconstructing the target object from the measured set of photoacoustic images is formulated as a linear inverse problem [Fig. 1(c)], studied at depth in many other instances of imaging in optics and other domains [8-12]. In principle, a reconstruction approach for solving such inverse problems should exploit all available information. This includes, in addition to the acquired images and detection system response, any prior information on the statistics or structure of the unknown illumination patterns, the non-negativity of the illumination intensity and the object absorption, and any inherent structural correlations or sparsity.

To this end, we employ a reconstruction approach based on compressed sensing (CS) [13-15]. CS has been demonstrated to enable super-resolved optical imaging of microscopic structures [11, 12, 16], and imaging using sub-Nyquist sampling [14, 17], i.e. imaging using a number of measurements that is lower than the number of image pixels. Our use of CS for super-resolved photoacoustics combines the high-resolution information contained in the temporal fluctuations, with the super-resolution recovery capability of CS, to retrieve the maximum amount of information using a minimum number of acquired photoacoustic frames. Unlike the use of dynamic speckle illumination for enhanced resolution and optical sectioning in optical microscopy [7, 8, 18, 19], [20-22], in photoacoustics the optical speckle grain size is not limited by the same diffraction-limit as the *imaging* point spread function (PSF), which is *acoustic*. In practice, the speckle grain size can be orders of magnitude smaller than the PSF dimensions [6]. This suggests that the resolution increase is not limited by the usual factor of two as in structured-illumination microscopy [7], even without nonlinearities in the imaging process [23]. Given the differences in dimensions, the speckle grains may be thought of as playing a similar role to blinking sources of signal ('molecules') with dimensions much smaller than that of the imaging system PSF. A situation analogous to the one considered in SOFI [5], PALM [24] and STORM [25] super-resolution microscopy techniques utilizing blinking fluorescent molecules.

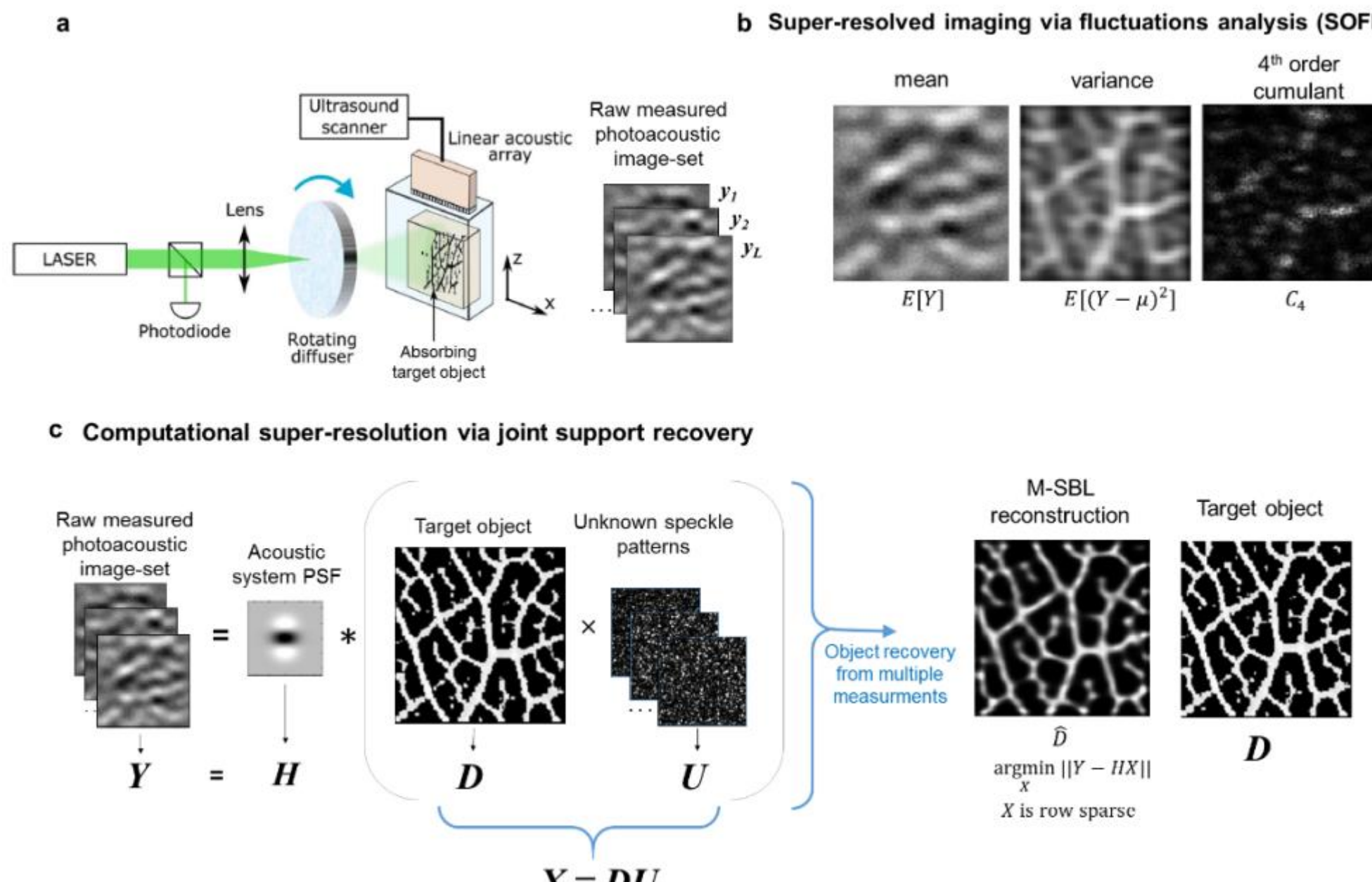

**Fig. 1. Concept and numerical example: (a)** Experimental setup. A pulsed laser beam is passed through a rotating diffuser producing temporally varying speckle patterns that illuminate the target. For each speckle pattern, ultrasound waves that are generated from the absorbing portions of the sample via the photoacoustic effect are recorded using a linear transducer array. The result is a set of photoacoustic images $Y=[y_1; y_2; \ldots; y_L]$ each with a resolution limited by the acoustic diffraction-limit. **(b)** An increase in resolution can be obtained by analyzing high-order statistical moments (cumulants) of the temporal fluctuations in the image-set via SOFI [4]. However, calculating higher order moments suffers from statistical noise due to the finite number of acquired images. **(c)** In the presented approach a compressed-sensing reconstruction algorithm takes advantage of available prior information on the object structure (here, sparsity of the absorbing structure, given by the diagonal matrix $D$), acoustic system response (given by the convolution matrix $H$), and optical speckle properties (the matrix $U$ whose columns are the individual speckle illumination patterns) to perform the recovery. The recovery is done over the matrix $X = DU$. The estimation of the object, $\hat{D}$, is then calculated as the row-wise standard deviation over $X$ (see text). The result is a reconstruction with improved fidelity and resolution. The reconstructions in (b-c) are numerical results obtained using 2000 speckle patterns with measurement noise of 5% of signal peak.

## 2. Principle

The principle of our approach is presented in Fig. 1(a,c). We consider the photoacoustic tomography experimental setup given in Fig. 1(a), employing an ultrasound transducer-array for detection, and a pulsed laser source with sufficiently large coherence length to produce speckles. A rotating diffuser in the illumination path produces random unknown optical speckle illumination patterns on the absorbing object that we wish to image. For each of the $m=1..L$ unknown speckle intensity patterns, given by $u_m(x,z)$, a single photoacoustic image, $y_m(x,z)$, is measured, where $x$ and $z$ are spatial coordinates (Fig. 1(a)). For simplicity of the analysis we consider a linear transducer array that images a small two-dimensional absorbing structure. The sought-after structure spatial absorption pattern is given by *o(x,z)*. We assume the object lies at the center of the transducer-array field of view (FOV), effectively having a shift-invariant acoustic PSF [4]. Under these considered assumptions, each acquired photoacoustic image, $y_m(x,z)$, is a convolution of the acoustic detection PSF, given by *h(x,z),* with the object structure, *o(x,z),* multiplied by the unknown speckle illumination pattern intensity $u_m(x,z)$:

$$y_m(x,z) = h(x,z) * [o(x,z) \times u_m(x,z)]. \quad (1)$$

The impulse response *h(x,z)* is constructed by back-propagation of the detected acoustic signals from a point absorber, in a prior calibration procedure (see Methods). In principle, the same analysis can be performed in three dimensions and for shift-varying PSF.

The imaging challenge is thus to find *o(x,z)* given the known/measured system PSF *h(x,z)*, and the set of photoacoustic images $y_m(x,z)$, without knowing the speckle patterns $u_x(x,z)$. Importantly, the photoacoustic images are of considerably lower resolution (lower spatial frequency bandwidth) than both the object and speckle patterns, due to the convolution with the acoustic PSF, *h(x,z)*. While in conventional structured illumination and in ghost-imaging [17] the speckle illumination patterns are known and the reconstruction is straightforward, here, as in blind-SIM, the speckle patterns are unknown. However, many of the speckle patterns properties, such as their non-negativity, intensity statistics and correlations, are universal [7] and can be used in the reconstruction algorithm. Importantly, since a speckle pattern is the result of random interference, the propagation through thick scattering tissue will only change the speckle realization but will not change the speckle characteristic properties and most importantly contrast, as long as the laser coherence length is sufficiently long. This important feature means that a high contrast and high resolution structured illumination is attained at depth, which would not be possible with incoherent illumination. While using SOFI to analyze the $N^{th}$-order statistical cumulant of $y_m$ yields, in principle, a $\sqrt{N}$ resolution increase without deconvolution [5], it is accompanied by strong artifacts when an insufficient number of frames is available [5] [Fig. 1(b), rightmost inset]. Since SOFI's statistical analysis does not take into account all available information besides the temporal fluctuations, its performance can be surpassed through a model-based approach that considers prior knowledge of the object and the system. This is exactly the design goal of CS: to recover the maximum amount of information from a minimal number of measurements. In a nutshell, a CS algorithm solves a set of underdetermined linear equations, such as those given by Eq. (1), by exploiting the inherent sparsity of natural objects in an appropriate transform basis. Remarkably, such sparsity is a general property of most natural images, and is at the core of modern lossy image compression algorithms, such as JPEG [14]. Here we exploit the sparsity of the object *O* in space.

To establish a CS framework for our problem, we formulate the problem that is given by Eq. (1) in a continuous coordinate space by its representation in an adequately sampled discrete space. The intensity pattern exciting photoacoustic signals is then given by a vector $u_m$ where its entries represent the intensity of the $m^{th}$ illumination pattern at all relevant spatial positions *(x,z)*. The acoustic spatial emission pattern is then written as $v_m = Du_m$, where $D = diag(O_1, \dots, O_N)$ is a non-negative diagonal matrix representing the object pattern on an *N* pixels grid. In the case where the object pattern is sparse in real space, *D* has a sparse diagonal. Each photoacoustic image $y_m$ is a result of a convolution of the photoacoustic emission pattern with the acoustic detection system PSF, given by the vector *h*. We denote by *H* its convolution matrix. With these notations the $m^{th}$ measured photoacoustic image can be written as [Fig. 1(c)]:

$$y_m = HDu_m. \quad (2)$$

Stacking the entire series of measurements $y_m$ as columns in a matrix *Y*, and $u_m$ as columns in a matrix U, the entire image set acquisition process may be described by [Fig. 1(c)]:

$$Y = HDU \quad (3)$$

where $Y$ and $H$ are the measured photoacoustic image-set and system response, correspondingly, and are assumed to be known. The matrices $D$ and $U$ are the unknown object absorption pattern (on the diagonal of $D$) and the unknown speckle illumination patterns (as columns of $U$).

Let $X=DU$. Since $D$ is diagonal and sparse, the support of $X$ and $D$ will be the same, where the support is defined as the nonzero rows of a matrix. Thus, we state our recovery problem as:

$$\underset{X}{\operatorname{argmin}} ||Y - HX||, \ \ s.t. \ \ x_{ij} \geq 0, X \text{ is row sparse.} \tag{4}$$

As the support of $X$ is equal to that of $D$, and since for fully developed speckles the ensemble average of the speckle intensity is the same for all spatial positions (meaning that the row-wise average value for $U$ is approximately the same for all of its rows), we calculate $\widehat{D}$ as the row-wise standard deviation over $X$ (See detailed explanation in Methods). The challenge of recovering a common support from the acquisition of multiple correlated sparse signals is referred to in the CS literature as the multiple measurement vector (MMV) problem [15, 26]. Several methods have been proposed for estimating the support [15]. Here we use a Bayesian approach to perform the recovery [8, 27, 28], referred to as multiple sparse Bayesian learning (M-SBL). We chose M-SBL since it led to superior reconstruction fidelity compared to other MMV methods we have tested. The M-SBL algorithm implements a maximum a-posteriori estimate (MAP) to find the optimal $X$ while defining some constraints on $X$ so as to encourage solutions which match the prior knowledge. In our setting, we employed a spatial sparsity prior (see details in Methods).

## 3. Results

### *3.1 Experimental results*

To experimentally demonstrate the advantage of the proposed reconstruction strategy over conventional photoacoustic reconstruction and statistical based fluctuations analysis, we used the above described M-SBL algorithm to analyze a set of experimental photoacoustic images of test samples made of absorbing beads of diameters 50µm-100µm, measured under dynamic speckle illumination [4]. The experiments were performed using the experimental system sketched in Fig. 1(a), and described in detail in [4] (see Methods).

The experimental results for three different samples are presented in Fig. 2. The leftmost column of Fig. 2 presents the optical image of the samples as imaged directly without the presence of any scattering. These can be considered as the 'ground truth' of *o(x,z)*. The other four columns in Fig.2 present images reconstructed from the photoacoustic acquired data in four different fashions: (1) a conventional photoacoustic image, given by the mean of the photoacoustic image set $y_{\text{conventional}} = \frac{1}{M}\sum_m y_m$ as obtained via back-propagation [29]; (2) a 2$^{nd}$ order SOFI fluctuation analysis followed by a Richardson-Lucy deconvolution using the squared PSF for deconvolution. These may be considered as the best results of photoacoustic fluctuations analysis, as achieved in [4]; (3) an M-SBL reconstruction using as input only the mean photoacoustic image, i.e. providing the increase in resolution relying only on sparsity without additional information from speckle fluctuations; and (4) an M-SBL reconstruction using the entire image dataset, providing the main result of this work. One may consider the results of SOFI as exploiting only the temporal fluctuations of the signal, the results for M-SBL on the standard photoacoustic image as optimal recovery using only the sparsity prior, and the rightmost column as exploiting *simultaneously* the sparsity priors, the common support and the fluctuations information for all of the speckle realizations. As expected, exploiting more information yields a superior reconstruction fidelity, recovering most accurately the number and positions of the absorbing beads, and reducing imaging artifacts. Since the PSF used to form the deconvolution matrix, $H$, was measured using beads of size similar to the imaged beads (see Methods), some of the reconstructed beads appear smaller than their real dimensions.

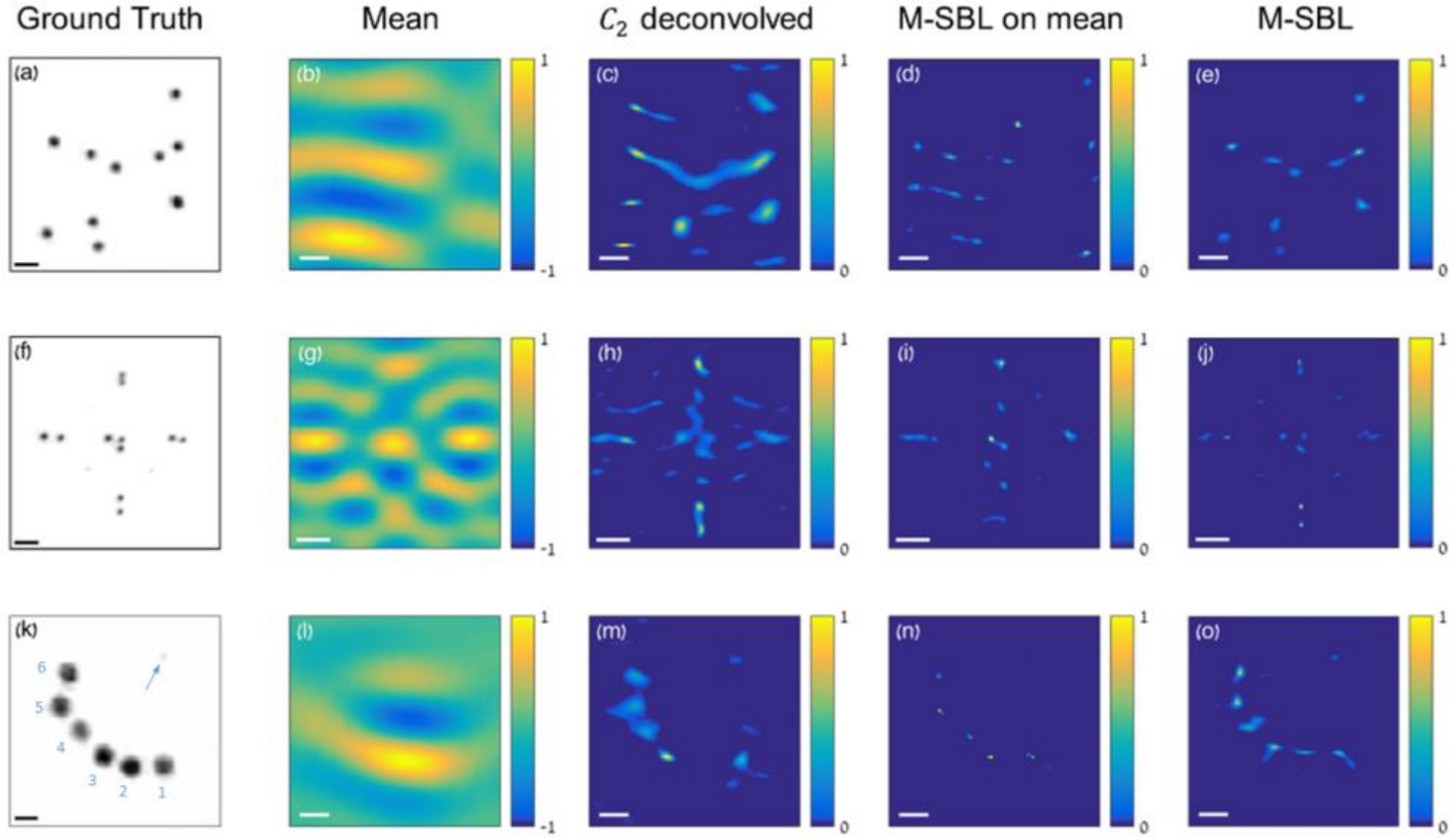

**Fig. 2.** Experimental comparison of reconstruction strategies: **(a,f,k)** Direct optical imaging of the absorbing beads without any scatterer. **(b,g,i)** conventional photoacoustic back-projection calculated by averaging all acquired photoacoustic images, **(c,h,m)** Reconstruction using 2nd-order SOFI fluctuation analysis followed by Richardson-Lucy deconvolution with the squared PSF, as presented in [4]. **(d,i,n)** M-SBL algorithm employing sparsity constraint ran using only the conventional photoacoustic image of (b,g,i). **(e,j,o)** M-SBL using all speckle illuminated images. Scalebars, 250 µm in (a-e), 400 µm in (f-j), 125 µm in (k-o). For M-SBL reconstructions, the variance of the recovered matrix $X$ is shown, for a fair comparison with SOFI 2nd order reconstruction.

### *3.2 Expected performance as a function of experimental parameters*

The resolution enhancement of the proposed nonlinear recovery approach depends on many experimental parameters. While it depends most critically on the size of the acoustic PSF (given laterally by the acoustic diffraction limit and axially by the transducer impulse response), it is also highly sensitive to the experimental signal to noise ratio (SNR), the absorbing sample structure/sparsity, and the speckle grain dimensions compared to the PSF dimensions. Qualitatively, the best performance is expected for the narrowest PSF, highest measurement SNR and the sparsest object.

To quantitatively analyze the expected reconstruction fidelity as a function of the above parameters we have numerically investigated a large set of imaging scenarios involving different PSF size, object sparsity, and SNR. The results of this study are presented in Fig. 3. Figure 3(a-b) display the correlation between the reconstructed images and the object for each of the considered scenarios. The vertical axis represents the sample sparsity/complexity, taken here as the number of absorbing pixels contained in the area enclosed by the PSF, where the pixel size is chosen as the optical diffraction limit (a speckle grain dimension). The horizontal axis provides a measure of the PSF size, taken as the ratio between the width of the PSF and the optical diffraction limit. All simulations were performed with a pixel grid of 70 by 70 pixels, with a pixel size equal to the speckle grain dimensions, i.e. no structures with dimensions below the optical diffraction-limit are considered. The PSF used in the simulations was generated by simulating the acoustic response of a 50um bead being uniformly illuminated by a 1ns laser pulse and recorded by a linear transducer array with 256 elements, upper frequency limit of 8MHz, inter-element pitch of 0.125mm, and element width of 0.125mm.

As expected from the intuitive qualitative description given above, both the PSF dimensions and the object sparsity play a crucial role in obtaining a high fidelity reconstruction. One can immediately appreciate that all of the results in the top row in the plots of Fig. 3(a-b) display a near perfect reconstruction. This perfect reconstruction results from the fact that the top row presents the case of very sparse samples, which contain nearly a single absorber inside a resolution cell given by the PSF dimensions. This is close to the scenario considered in localization microscopy techniques such as PALM/STORM, where a centroid analysis may provide a good estimation of the absorber location under a sufficiently low labelling density. It is important to note that while in SOFI, PALM and STORM the labelling density may be controlled by the labeling concentration and excitation laser power, this is not possible in linear photoacoustic imaging of endogenous contrast.

To illustrate what recovery errors appear when the reconstructed images fail to perfectly recover the object structure, we provide in Fig. 3(l-o) several examples for reconstructed images side by side with the original objects (h-k), for several cases presented in Fig. 3(b). It can be seen that when the concentration of absorbers is too high (i.e. sparsity is too low) the algorithm fails to identify their exact individual positions but instead delivers some continuous line-like structure connecting them, which can be interpreted as blurring of the original image. This gradual 'break-down' of the recovery algorithm is encouraging for practical imaging purposes, as even the reconstructions with low calculated correlation to the original object carry relevant information on the object. This may be advantageous when continuous structures such as blood vessels are considered. While different measures for the sample sparsity, SNR, and PSF size relative to the object structure may be chosen, we expect the results presented in Fig. 3(a-b) to serve as a reference to the expected performance given a specific experimental imaging scenario.

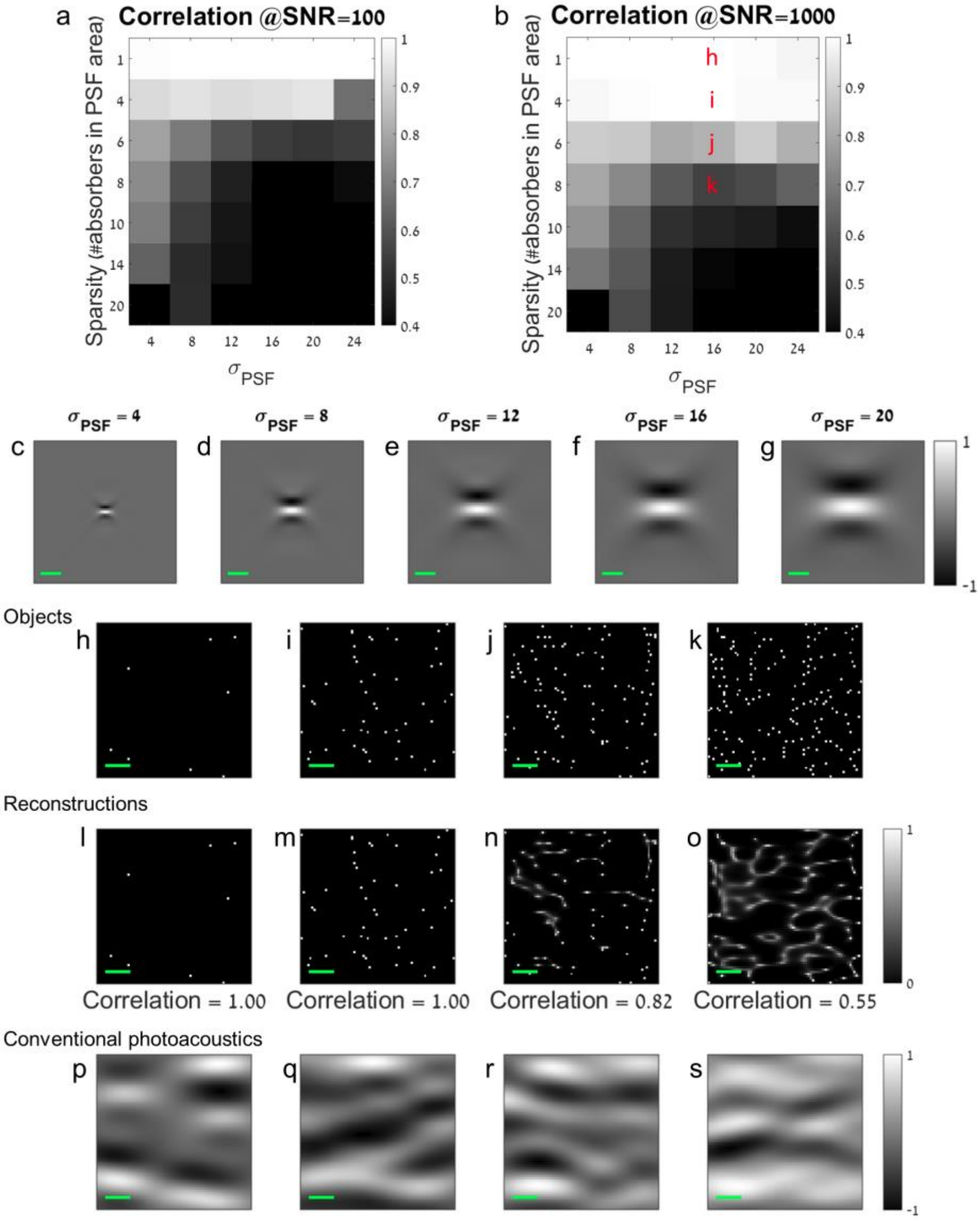

**Fig. 3.** Numerical study of the M-SBL reconstruction fidelity as a function of SNR, PSF dimensions, and sparsity. **(a-b)** Correlation between the simulated objects and the reconstruction. Horizontal axis: width of the acoustic PSF (in pixels, pixel size = speckle grain dimensions); vertical axis: sample sparsity, taken here as the number of absorbing pixels contained in the area enclosed by the PSF. Note the gradual transition between success (high correlation) and failure. **(c-g)** The different PSFs used for the simulations of (a-b), the $\sigma_{PSF}$ is measured as the full width at 77% of the max, **(h-k)** Examples of objects used to obtain four of the points in (b), **(l-o)** M-SBL reconstruction of the corresponding objects and their correlations with the object pattern. Reconstructions shown are the standard deviation of X. **(p-s)** The conventional photoacoustic images (mean of acquired image set).

## 4. Discussion

We proposed an advanced reconstruction algorithm for photoacoustic imaging, which efficiently exploits dynamic temporal fluctuations, joint sparse support constraints, and a known system response for improved resolution and reconstruction fidelity. Our approach provides superior performance compared to the recently proposed SOFI-based photoacoustic speckle fluctuation analysis [4].

From an estimation theory point-of-view, it is clear that improved performance will be obtained for a reconstruction algorithm that takes into account all available information. Here, we used CS to take into account some of this information by exploiting the object sparsity and the multiple random projections provided by the random speckle illumination. Improved algorithms could be developed by incorporating also the non-negativity of the speckle intensity patterns and object structure, and the known universal Rayleigh statistics of fully developed speckles [30], which we have used only implicitly here to reconstruct the object from the reconstructed matrix $X$ (see Methods).

We have demonstrated our approach using two-dimensional objects and sparsity constraints in real space, however one may consider applying the proposed CS recovery approach to reconstruct three-dimensional (3D) objects utilizing sparsity in any other 3D-sparse transform basis representations (or 4D, for dynamic objects), which better matches the object structure, e.g. wavelet or minimum total variance (min-TV).

We formulated here the recovery problem in image space (x,z), although the raw measured acoustic signals are time-domain signals, i.e., the convolution matrix $H$ in equations (2-3) was taken as a combination of acoustic forward mapping and subsequent computational back-projection. The recovery problem may as well be formulated in the raw acoustic data space, i.e., taking $H$ as the acoustic forward operator and $y$ as the raw acoustic data. This is an interesting point, which is currently under investigation. It may have possible advantages over the image-space formulation, such as the insensitivity to the specific back-propagation algorithm used.

In this work we made use of the measured system PSF. However, when the system PSF is not measured it may be possible to develop an advanced reconstruction algorithm that estimates the PSF and the object simultaneously, as is done in blind deconvolution [31] and blind compressed sensing [32].

An important practical challenge for applying the approach for deep tissue photoacoustic imaging arises from the large difference between the speckle grain dimensions and the acoustic PSF dimensions. At large imaging depths the speckle grain dimensions, which are given by the optical diffraction limit, would be orders of magnitude smaller than the imaging acoustic PSF [6]. In this scenario, the measured value at each pixel in the raw photoacoustic frames is the result of a sum over a large number of fluctuating uncorrelated speckle grains (the PSF convolution kernel being much larger than the speckle grain). This results in an overall small fluctuation to mean value in each image pixel between the different frames [6]. Resolving the small fluctuations over a large background may be challenging under low SNR conditions. Choosing a long optical wavelength and a high ultrasound frequency would be advantageous for this task.

We formulated the photoacoustic imaging problem in the case of dynamic speckle illumination as an instance of blind structured illumination (blind-SIM), and experimentally demonstrated it on two-dimensional objects using a transducer array. As such, other algorithms solving the blind-SIM problem could be employed and their performance compared to the specific algorithm used here. For example, during the final preparations of this manuscript a blind-SIM photoacoustic imaging of a one-dimensional object was demonstrated [33].

## 5. Methods

### *5.1 Experimental setup*

The setup used to perform the experiments is drawn schematically in Fig. 1. The beam of a nanosecond pulsed laser (Continuum Surelite II-10, 532 nm wavelength, 5 ns pulse duration, 10 Hz repetition rate) was focused on a ground glass diffuser (Thorlabs, 220 grit, no significant ballistic transmission). The scattered light illuminated a 2-D absorbing sample embedded in an agarose gel block. This phantom was located 5 cm away from the diffuser, leading to a measured speckle grain size of ~30 µm. The absorbing sample was placed in the imaging plane of a linear ultrasound array (Vermon, 4 MHz center frequency, >60% bandwidth, 128 elements, 0.33 mm pitch), connected to an ultrasound scanner (Aixplorer, Supersonic Imagine, 128-channel simultaneous acquisition at 60 MS/s). A collection of black polyethylene microspheres (Cospheric, 50 µm and 100 µm in diameter) was used to fabricate phantoms with isotropic emitters. The PSF of the system was measured for each sample by concentrating light on one single 50 µm-diameter bead. The diffuser was removed from the light path during this step, to ensure a homogenous illumination of the bead.

For each sample, a set of photoacoustic signals for 100 uncorrelated speckle patterns was obtained by rotating the diffuser. Special care was taken to reduce sources of fluctuations other than the multiple speckle illumination between photoacoustic acquisitions. The raw recorded (RF) acoustic signals were processed by a low-pass filter with a sharp cutoff that eliminated all frequencies above 2.4MHz in sample 1, and 2.8MHz and 5.3MHz for samples 2 and 3 correspondingly. Different cutoff frequencies were used to create more challenging recovery scenarios for the different algorithms. The dimensions of the resulting PSF, defined as the full width at half maximum (FWHM) after the low-pass filtering were 613/1643um, 537/1231um and 393/562um in the transverse/axial directions for samples 1, 2 and 3 correspondingly.

For each sample, 100 photoacoustic images were reconstructed from the raw acoustic signals, for each of the 100 speckle patterns, using a time-domain backprojection algorithm on a grid of 814 by 814 pixels with a pixel pitch of 25um. The time domain backprojection reconstruction is based on summing the photoacoustic signals taken at appropriate retarded times [29]. The reconstructed photoacoustic images $y_m$ were downsampled to half of their original size by bilinear interpolation, to reduce the required computational resources and run time. The total run time for a single reconstruction on a machine with an Intel i7 3.60GHz processor with 4 cores is about 10 minutes.

### *5.2 Recovery algorithm*

The MMV recovery algorithm we used in this work is M-SBL [27, 28]. The algorithm source code is available for download at [34]. As mentioned above, the M-SBL algorithm implements a maximum a posteriori estimator. Through the application of the Bayes law it searches for the value of $X$ which maximizes the joint probability of $P(X, Y)$, as detailed below. The M-SBL algorithm we adapted in this work assumes fluctuations having a Gaussian statistical distribution with zero mean [28]. Dynamic speckle illumination having a speckle grain size that is considerably smaller than the reconstruction grid indeed provides a Gaussian statistical distribution for the temporal fluctuations (as a direct consequence of the central limit theorem and the large number of summed speckle grains in each pixel), but with a mean that is not zero. Thus, when running M-SBL for the MMV case, the pixel-wise calculated temporal mean of the fluctuation images was subtracted pixel-wise. To provide a fair comparison with the 2$^{nd}$ order SOFI reconstruction, the M-SBL reconstructions presented in Figures 1-2 show the variance over each row of the recovered matrix $X$ which corresponds to the temporal variance at each reconstructed image pixel. In Figure 3 the displayed M-SBL results are the standard deviation of each reconstructed image pixel, since the standard deviation provides a measure that is linearly related to the mean absorption in each spatial position (pixel). In the case where the reconstruction grid pixel size is smaller than the speckle grain dimensions, the standard deviation of each pixel provides a quantitative estimate of the mean absorption since for the exponential statistics of fully developed speckle [30] the temporal standard deviation is equal to the mean. Since the speckles fluctuations are uncorrelated, in the case where the grid pixel size is larger than the speckle grain dimensions, the standard deviation provides the mean absorption times the square root of the number of speckles contained in the absorbing area in the pixel.

The M-SBL method used in this work is an adaptation of the algorithm of Zhang et al. [28]. Briefly, under certain prior assumptions of Gaussian distributions on the signal and noise, the posterior density of the *j*-th column of X becomes [27]:

$$p\left(X_{.j} \middle| Y_{.j};\gamma_j\right) = N\left(\mu_{.j},\Sigma\right), \tag{5}$$

where *X* and *Y* are the matrix to be reconstructed and the measurement matrix, correspondingly, $N(\mu_{.j},\Sigma)$ is the normal probability distribution with vector mean $\mu_{.j}$ and covariance matrix $\Sigma$, and $\gamma_j$ is an unknown variance hyperparameter of the *i*th row with no assumed prior: $p(X_{i.};\gamma_i) = N(0,\gamma_i I)$. The mean and covariance of (5) are given by [27]:

$$[\mu_{.1},\dots,\mu_{.L}] = E\{X|Y;\gamma\} = \Gamma H^T \Sigma_y^{-1} Y \tag{6}$$

$$\Sigma = \Gamma - \Gamma \mathrm{H}^T \Sigma_y^{-1} H \Gamma, \tag{7}$$

where $\Gamma = \mathrm{diag}(\gamma)$, $\Sigma_y = \sigma^2 I + H\Gamma H^T$ and $\sigma^2$ is the noise variance.

The values of $\gamma$ represent the prior distribution underlying the generation of the data *Y*. As stated by Wipf [27] the M-SBL can be seen as maximizing the cost function:

$$\mathcal{L}^\gamma(\gamma) = \sum_{j=1}^{L} t_{.j}^T \Sigma_y t_{.j} + L\log|\Sigma_y|. \tag{8}$$

The maximization is performed by equating the derivative with respect to $\gamma$ to zero, which results in the following update rule for $\gamma$ [27]:

$$\gamma_{i+1} = \frac{\frac{1}{L}\|\mu_{i.}\|_2^2}{1-\gamma_i^{-1}\Sigma_{ii}}, \qquad i = 1,..,k. \tag{9}$$

The final algorithm consists of the following steps [28]:

1. Initialize $\gamma_i = 1$, for all *i*
2. Calculate the values of $\Sigma$ and $\mu$ according to Eq. (6) and Eq. (7)
3. Update $\gamma$ using Eq. (9). If $1-\gamma_i^{-1}\Sigma_{ii}$=0 then no update is done for the corresponding pixels
4. Iterate over steps 2) and 3) for a fixed number of times k, in our case we used k=20
5. To estimate the original object $\widehat{D}$ we used the standard deviation of rows of $\mu$:

$$\widehat{D} = \begin{bmatrix} \sqrt{\sum_{t=1}^{L}\left|\mu_{1t}^{(k)}\right|^2} \\ \sqrt{\sum_{t=1}^{L}\left|\mu_{2t}^{(k)}\right|^2} \\ \vdots \\ \sqrt{\sum_{t=1}^{L}\left|\mu_{nt}^{(k)}\right|^2} \end{bmatrix}. \tag{10}$$

In case the signal-to-noise ratio is known it can be used to choose $\sigma$. In our work we chose values of $\sigma$ in the range of 0.1 to 0.5 that were found to give good reconstruction on our experimental data.

The source code of the algorithm used in this work is available for download at [34].

## Funding

This project has received funding from the European Research Council (ERC) under the European Union's Horizon 2020 research and innovation program (grants no. 278025, 677909, 681514), and Human Frontiers Science Program. O.K. was supported by an Azrieli Faculty Fellowship. S.G. acknowledge support from the institut Universitaire de France. E.B. acknowledges support from the Ecole Polytechnique Fédérale de Lausanne (EPFL) via a Visiting Professor Fellowship.

## Acknowledgment

We thank Dr. Zhilin Zhang for providing the basis for the M-SBL code.